\documentclass[twocolumn,10pt,5p]{elsarticle}

\usepackage{atlasphysics} 

\usepackage{subfigure}
\usepackage{mathrsfs}
\usepackage{graphicx}
\usepackage{multirow}
\usepackage{rotating}
\usepackage{xspace}
\usepackage{hyperref}

\newcommand{\RP}{\ensuremath{R}-parity}

\def\lumi{33~\ipb}

\def\taumc{\ensuremath{\tau_{\rm MC}}}

\def\rdv{\ensuremath{r_{\rm DV}}}
\def\zdv{\ensuremath{z_{\rm DV}}}
\def\mdv{\ensuremath{m_{\rm DV}}}
\def\ntdv{\ensuremath{N^{\rm trk}_{\rm DV}}}

\def\zpv{\ensuremath{z_{\rm PV}}}

\def\mm{\rm mm}

\def\purity{\ensuremath{R_{\rm int}}}

\def\beq{\begin{equation}}
\def\eeq{\end{equation}}
\def\beqa{\begin{eqnarray}}
\def\eeqa{\end{eqnarray}}

\journal{Physics Letters B}

\begin{document}

\begin{frontmatter}


\title{\vspace{-0.5cm} \flushright{\normalsize{CERN-PH-EP-2011-131}} \\
\center{\bf \boldmath Search for displaced vertices arising from decays of new heavy particles in 7~TeV $pp$ collisions at ATLAS}}

\author{The ATLAS Collaboration}


\begin{abstract}

We present the results of a search for new, heavy particles that decay at a
significant distance from their production point into a final state
containing charged hadrons in association with a high-momentum muon.
The search is conducted in a $pp$-collision data sample with
a center-of-mass energy of 7~\tev\ and
an integrated luminosity of $\lumi$ collected in 2010 by the ATLAS
detector operating at the Large Hadron Collider.
Production of such particles is expected in various
scenarios of physics beyond the standard model.
We observe no signal and place limits on the production cross-section 
of supersymmetric particles in an R-parity-violating scenario
as a function
of the neutralino lifetime. Limits are presented for different
squark and neutralino masses, enabling extension of the limits to 
a variety of other models. 

\end{abstract}

\end{frontmatter}


%
%
\section{Introduction}
\label{sec:Introduction}

Various scenarios of physics beyond the standard model
predict the production at the Large
Hadron Collider (LHC) of heavy particles with lifetimes that may be of order
picoseconds to about a nanosecond. 
An example of such a scenario is gravity-mediated supersymmetry (SUGRA) with
\RP\ violation (RPV), where current limits on RPV
couplings~\cite{Allanach:2006st} allow for the decay vertex of the
lightest supersymmetric particle to be within the range
accessible to collider-based particle detectors.
In gauge-mediated supersymmetry models, the 
next-to-lightest supersymmetric particle may be long lived due to  
suppression of its decay by the large
supersymmetry-breaking scale~\cite{Dimopoulos:1996vz}.
Additional scenarios allowing for such a signature include 
split-supersymmetry~\cite{Hewett:2004nw}, 
hidden-valley~\cite{Strassler:2006im}, 
dark-sector gauge bosons~\cite{Schuster:2009au}, 
stealth supersymmetry~\cite{Fan:2011yu},
or a meta-stable supersymmetry-breaking sector~\cite{Zur:2008zg}. 

Searches for related signatures have been performed at the Tevatron
with $\sqrt{s}=1.96~\tev$ $p\bar p$ collisions. The D0 collaboration
has searched for a long-lived neutral particle decaying into a final
state containing two muons~\cite{D0:2006} or a $b\bar b$
pair~\cite{D0:2009}.  No signal was observed, and limits were computed
in the context of RPV and hidden-valley model scenarios.

\begin{figure}[!hbtp]
   \centering \includegraphics[width=5cm]{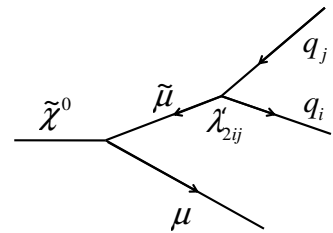}
   \caption{\label{fig:RPV-diagram}
    Example of a diagram of a new massive particle $\tilde\chi^0$
    (such as the lightest neutralino) decaying into
   a muon and two jets via a virtual smuon, 
   with lepton-number and $R$-parity violating coupling $\lambda'_{2ij}$.}
\end{figure}

In this letter, we report the results of a search for a heavy particle
decaying into several charged particles at a distance of order
millimeters to tens of centimeters from the $pp$ interaction point,
in events containing a muon with high transverse momentum (\pt).  
We report the results of the search in terms of limits within
the SUGRA scenario, where this signature corresponds to the
decay of the lightest supersymmetric particle
due to non-zero RPV couplings $\lambda'_{2ij}$, via a diagram
such as the one shown in Fig.~\ref{fig:RPV-diagram}. However, it may
also be the result of other models with heavy, long-lived particles
that decay into or are produced in association with a high-$\pt$ muon.

\section{The ATLAS detector}
\label{sec:detector}

The ATLAS detector~\cite{ATLAS:detector2008} comprises a tracking
inner detector (ID) system, a calorimeter system, and an extensive
muon spectrometer (MS). 

The ID operates in a 2~T magnetic field and provides
tracking and vertex information for charged particles in the
pseudorapidity range $|\eta| < 2.5$, where $\eta \equiv
-\ln\tan(\theta/2)$ and $\theta$ is the polar angle, defined with
respect to the cylindrical symmetry axis (the $z$ axis) of the detector.
At small radii, high-resolution pattern recognition capability is
obtained using silicon pixel layers and stereo pairs of silicon
microstrip layers.  
The pixel system comprises three barrel layers, and three forward
disks on each side of the interaction point.  
Of particular significance to this analysis are the barrel pixel
layers, which are positioned at radii of 50.5, 88.5, and 122.5~\mm.
The silicon microstrip
tracker (SCT) has four barrel layers, and nine forward disks on each
side.
At larger radii, a transition-radiation tracker (TRT) composed of
straw-tube elements interleaved with transition-radiation material
contributes to track reconstruction up to $|\eta| = 2.0$ and improves
electron identification.

The calorimeter system covers the pseudorapidity range $|\eta| <
4.9$. It includes a lead/liquid-argon electromagnetic calorimeter and
a hadronic calorimeter comprising a steel/scintillating-tile barrel
system and a liquid-argon endcap system with copper and tungsten absorbers.

The MS provides muon identification and momentum measurement in the
pseudorapidity range $|\eta|<2.7$ using a 3-layer system of gas-filled
precision-tracking chambers. The $|\eta|<2.4$ range is covered by
separate trigger chambers, which are used for the level-1 hardware
trigger. The MS operates in a toroidal magnetic field generated by a
barrel toroid and two endcap toroids.

A three-level trigger system is used for online event selection.  
It comprises a hardware-based level-1 trigger, which uses
information from the MS trigger chambers and the calorimeters, 
followed by two software-based trigger levels.

\section{Data and simulation}
\label{sec:data}

The data used in this analysis were collected between June and
October 2010.  Basic requirements on stable beam, detector, and trigger
conditions are applied, resulting
in a total integrated luminosity of \lumi.
%

Monte Carlo (MC) simulated event samples are used to study possible
sources of background, evaluate acceptance and signal efficiency,
and for systematic-uncertainty studies. 
For this purpose, we use minimum-bias, QCD-dijet, 
$W +$~jets, and $Z+$~jets samples produced with the
PYTHIA~\cite{pythia} MC generator, and $t\bar t$ samples generated
with MC@NLO~\cite{MC_NLO}.

We use PYTHIA to generate signal events, where a pair of primary
squarks or a squark-antisquark pair 
is produced in the $pp$ collision, and each squark (antisquark) decays into a
neutralino and a quark (antiquark). 
The signal events are generated with a gluino mass of 5~\tev, so
that the cross-section for gluino-pair production is small, and this 
process is not considered.

The parameters of the different signal MC samples are shown in
Table~\ref{tab:sigMC}. 
The generated values of squark and neutralino masses 
are chosen so as to provide
a broad range of neutralino velocities and daughter-particle
multiplicities, which are the quantities that have the greatest impact
on the signal efficiency (see discussion in 
Section~\ref{sec:eff})\footnote{The final neutralino mass values are the 
result of Higgsino-gaugino mixing.}.
The signal MC samples are labeled in Table~\ref{tab:sigMC} according to 
the squark mass and neutralino
mass, respectively: MH (medium-mass squark, heavy neutralino),
ML (medium-mass squark, light neutralino),
LL (light squark and light neutralino), and 
HH (heavy squark and heavy neutralino).
The RPV coupling $\lambda'_{211}$ is set to a non-zero value, so that
each neutralino decays to $\mu^- u \bar d$ (or the charge-conjugate
state).  The value of $\lambda'_{211}$ for each sample is chosen so
that a significant fraction of neutralino decays takes place
throughout the pixel region, and the resulting average neutralino
proper flight distances ($c\taumc$, where $\taumc$ is the generated 
neutralino lifetime) are given in Table~\ref{tab:sigMC}.
We note that while the MC distributions shown in Figs.~\ref{fig:eff},
\ref{fig:eff-2d}, and \ref{fig:massVsMult}
depend on $c\taumc$, the results of the analysis, presented in
Section~\ref{sec:results}, are independent of the values of
$\lambda'_{211}$ and $c\taumc$ with which the signal MC samples are
generated.

The signal cross-section values listed in Table~\ref{tab:sigMC} are
obtained from PROSPINO~\cite{Beenakker:1996ed}, which performs a
next-to-leading-order calculation. We observe that the cross-sections
for $\tilde q \tilde q$ and $\tilde{q} \bar{\tilde{q}}$ events are
significantly different between PYTHIA and PROSPINO, and that the
neutralino-velocity distributions generated by PYTHIA for these two
types of events are different, especially in the 150-\gev\ squark
case. To take this into account, $\tilde q \tilde q$ and $\tilde q
\bar{\tilde{q}}$ events signal samples are generated separately 
with PYTHIA, and reweighted according to the relative cross-sections
of the two event types in PROSPINO.

\begin{table}
\begin{center}
\renewcommand\arraystretch{1.1}
\begin{tabular}{|c|c|c|c|c|c|}
\hline
Sample & $m_{\tilde{q}}$  & $\sigma$
       & $m_{\tilde{\chi}^0_1}$   
       &  $\left<\gamma\beta \right>_{\tilde{\chi}^0_1}$  
       & $c\taumc$ \cr
    &  [GeV]           & [fb]     & [GeV]  &  & [\mm] \cr
\hline
MH & 700  & 66.4               & 494   & 1.0  &  78 \\
ML & 700  & 66.4               & 108   & 3.1  &  101 \\
LL & 150  & $539\times 10^3$   & 108   & 1.5  &  196 \\
HH & 1500 & 0.2                & 494   & 1.9  &  82 \\
\hline
\end{tabular}
\caption{\label{tab:sigMC}
Parameters of the four signal MC samples used in the analysis: the
squark mass, 
production cross-section from PROSPINO~\cite{Beenakker:1996ed}, 
neutralino mass, 
average neutralino boost factor from PYTHIA~\cite{pythia},
and average proper flight distance.  }
\end{center}
\end{table}

The background and signal MC samples are generated with an average of
$\left<\mu \right> = 2.2$ interactions per
event, corresponding to $\left< \mu \right>$ for data taken in 2010. 
An additional sample, otherwise identical to the
MH sample in Table~\ref{tab:sigMC}, 
is produced with $\left< \mu \right> = 5$.
Each simulated event is passed through the GEANT4-based~\cite{geant4}
ATLAS detector simulation~\cite{ATLASgeant4} and processed in the same
way as the data.

\section{Vertex reconstruction and event selection}
\label{sec:cuts}

We select events that pass an MS level-1 single-muon trigger, where
the muon candidate has transverse momentum $\pt > 40$~\gev, as
determined in the MS. 
We require the primary vertex (PV) originating from the
$pp$ collision to have at least five tracks and a $z$ position in the range
$|\zpv|<200~\mm$. In events with multiple interactions, the PV with
the highest scalar sum of the $\pt$ of its tracks is used.

Reconstruction of a displaced vertex (DV) begins with the selection of
tracks with $\pt > 1$~\gev\ formed with at least two SCT hits.
For each track we require $|d_0|>2~\mm$, where $|d_0|$ is the 
impact parameter of the track 
with respect to the transverse position of the PV,
$(x_{\rm PV}, y_{\rm PV})$. In the MC, this requirement rejects 98\% of all 
tracks originating from the primary $pp$ interaction.

The selected tracks are used to search for displaced vertices using an
algorithm based on the incompatibility-graph approach, 
similar to that used in Ref.~\cite{vtxNote}.
The algorithm begins by reconstructing 2-track seed vertices
from all track pairs,
and keeping those that have a vertex-fit $\chi^2$ less than 5.
A seed vertex is rejected if one of its tracks has hits between the
vertex and the PV.
Seed vertices are combined into multi-track vertices in an iterative
process, as follows.  
If a track is used in two different vertices, the action taken depends
on the distance $D$ between the vertices: if $D < 3 \sigma_D$, where
$\sigma_D$ is the estimated uncertainty on $D$, then the tracks of the
two vertices are combined and refitted to a single vertex; otherwise,
the track is associated with only the vertex relative to which it has
the smaller $\chi^2$.
If the $\chi^2$ of a track relative to the resulting vertex is greater
than 6, the track is removed from the vertex, and the vertex is
refitted. The process continues until no tracks are shared among
different vertices.
Finally, vertices that are separated by less than 1~\mm\ are combined
and refitted.
Events containing at least one such displaced vertex are said to
satisfy the event selection criteria.

The typical position resolution of the DV in the signal MC samples is
tens of microns for \rdv\ and about 200 microns for \zdv\ near the
interaction point. For vertices beyond the outermost pixel layer,
which is located at a radius of 122.5~\mm, the typical resolution is
several hundred microns for both coordinates.

To ensure the quality of the DV fit, we require the $\chi^2$ per
degree of freedom (DOF) of the fit to be less than 5.
The DV position is required to be within the barrel pixel fiducial
region, defined by the longitudinal and transverse ranges
$|\zdv|<300$~mm, $\rdv<180$~mm, respectively.
To suppress background from tracks that originate from the PV, we require the
transverse distance 
$\sqrt{(x_{\rm DV} - x_{\rm PV})^2 + (y_{\rm DV} - y_{\rm PV})^2}$
between the primary and
the displaced vertices to be at least 4~mm.
We require the number of tracks \ntdv\ in the DV to be at least four, to
suppress background from random combinations of tracks and from
material interactions.
Background due to particle interactions with material is further suppressed by
requiring $\mdv > 10$~\gev, where \mdv\ is the invariant mass of the
tracks originating from the DV.
We refer to vertex candidates that satisfy (fail) the $\mdv > 10$~\gev\
requirment as high-\mdv\ (low-\mdv) vertices.

\begin{figure}[!hbtp]
\begin{center}
\includegraphics[width=8cm]{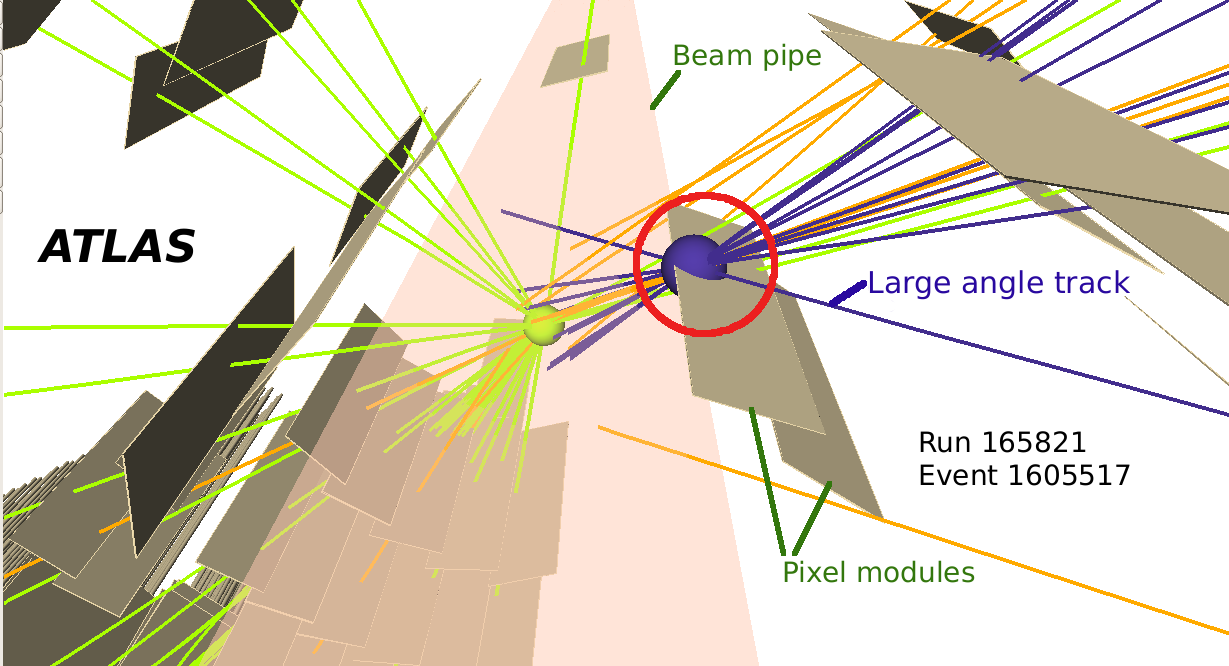}
\caption{\label{fig:high-mass-bgd-vtx} An event from a jet-trigger
data sample, where a high-mass vertex (circled) is the result of an
apparently random, large-angle intersection between a track 
(labeled as ``Large angle track'') and a
low-\mdv\ hadronic-interaction vertex produced in a pixel module. 
Tracks originating from this vertex are shown in blue, those
from the primary vertex are green, and other tracks are orange.
The beampipe and pixel modules with track hits are shown.
}
\end{center}
\end{figure}

Low-\mdv\ vertices from particle-material interactions are abundant in
regions of high-density detector material.  High-\mdv\ background may
arise from random spatial coincidence of such a vertex with a
high-\pt\ track, especially when this track and the particle that
created the material-interaction vertex originate from different
primary interactions, which may result in a large angle between their
momentum vectors. An example of such a random combination of a
material-interaction vertex with a high-$p_T$ track is shown in
Fig.~\ref{fig:high-mass-bgd-vtx}.

To suppress this type of background, we veto vertices that are
reconstructed within regions of high-density material, mapped using
low-\mdv\ material-interaction candidate vertices in data and 
true material-interaction vertices in minimum-bias MC events.
We use the \zdv\ and \rdv\ positions of these vertices to
form a 2-dimensional material-density map with a bin size of
4~mm in \zdv\ and 1~mm in \rdv. 
Studies have shown~\cite{vtxNote} that the positions of pixel
layers and associated material are well simulated in the 
MC detector model, while the simulated beampipe position is shifted
with respect to the actual position. 
Thus, the use of data events to construct the material map ensures the
correct mapping of the beampipe material, while MC events
make possible the high granularity of the map at the outer
pixel layers, where material-interaction vertices
in the data are relatively rare due to the low density of 
primary particles.
Material-map bins with vertex density greater than an \rdv- and
\zdv-dependent density criterion are designated as
high-density-material regions, which constitute 34.4\% of the fiducial
volume $|\zdv|<300$~mm, $4<\rdv<180$~mm.
High-\mdv\ vertices reconstructed within these bins are rejected.
We refer to the combination of all the 
requirements above as the vertex-selection criteria.

In addition to the vertex-selection criteria,
events are required to contain a muon candidate
reconstructed in both the MS and the ID with $\pt>45~\gev$, which is
well into the efficiency plateau of the $40~\gev$ level-1 trigger.
The muon candidate must satisfy $\sqrt{\Delta\phi^2 + \Delta\eta^2} <
0.1$, where $\Delta\phi$ ($\Delta\eta$) is the difference between the
azimuthal angle (pseudorapidity) of the reconstructed muon candidate
and that of the muon identified by the trigger.
The ID track associated with the muon candidate is required to have at
least six SCT hits, with at most one SCT hit that is expected but not
found, and must satisfy an $|\eta|$-dependent requirement on the
number of TRT hits. No pixel-hit requirements are applied to the muon
track. The muon track is not required to originate from a DV, in order
to maintain high signal efficiency and sensitivity to scenarios where
the muon may originate from the PV rather than from the DV, such
as supersymmetry with longer decay chains involving prompt smuons. 
The combination of requirements above is referred to as the
muon-selection criteria.

\section{Signal efficiency}
\label{sec:eff}

\begin{figure}[!hbtp]
   \centering \includegraphics[width=9cm]{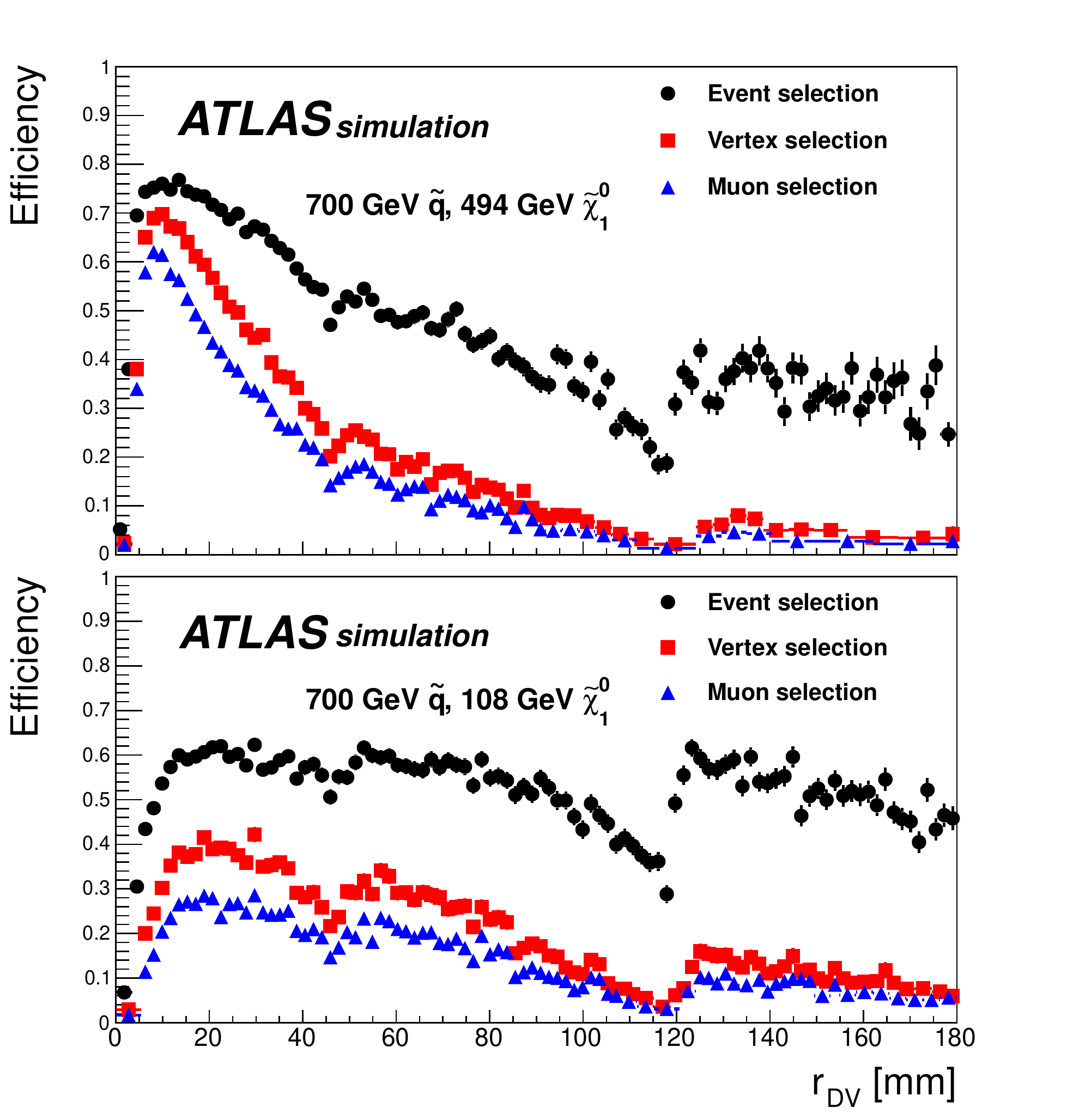}
   \caption{\label{fig:eff} The efficiency as a function of \rdv\ for
     vertices in the signal MC samples~MH (top) and~ML (bottom) of
     Table~\ref{tab:sigMC}.  In each plot, the top histogram (circles)
     is the efficiency for the event selection criteria, the middle
     histogram (squares) shows the efficiency when also requiring one
     vertex to satisfy the vertex-selection criteria, and the bottom
     histogram (triangles) includes the muon-selection criteria.  The
     material veto is not applied when making these plots.}
\end{figure}

The efficiency for signal MC vertices is shown in Fig.~\ref{fig:eff} as
a function of \rdv\ for the different selection criteria and two
representative signal MC samples -- sample~MH of Table~\ref{tab:sigMC}, 
where the neutralino is heavy
and slow moving, and sample~ML, where it is light and highly boosted.  
The efficiency presented in Fig.~\ref{fig:eff} depends strongly on the 
efficiencies for track reconstruction and track selection, which are 
affected by several factors:
(1) The number of tracks originating from the DV and their total invariant
mass increase with the neutralino mass.
(2) More tracks fail the $|d_0|>2$~mm requirement for small \rdv\ or
if the neutralino is highly boosted.
(3) Track-finding algorithms used in track reconstruction are not
optimized for tracks with large values of $|d_0|$, resulting in
decreased efficiency for large \rdv.
(4) One of the track-quality requirements applied by standard ID
tracking is that tracks share no more than one hit in the pixel and
SCT layers. As a result, the efficiency is low when \rdv\ is slightly
smaller than the radial position $r_{\rm pixel}$ of a pixel layer and
increases again for $\rdv > r_{\rm pixel}$. 

\begin{figure}[!hbtp]
   \centering 
   \includegraphics[width=9cm]{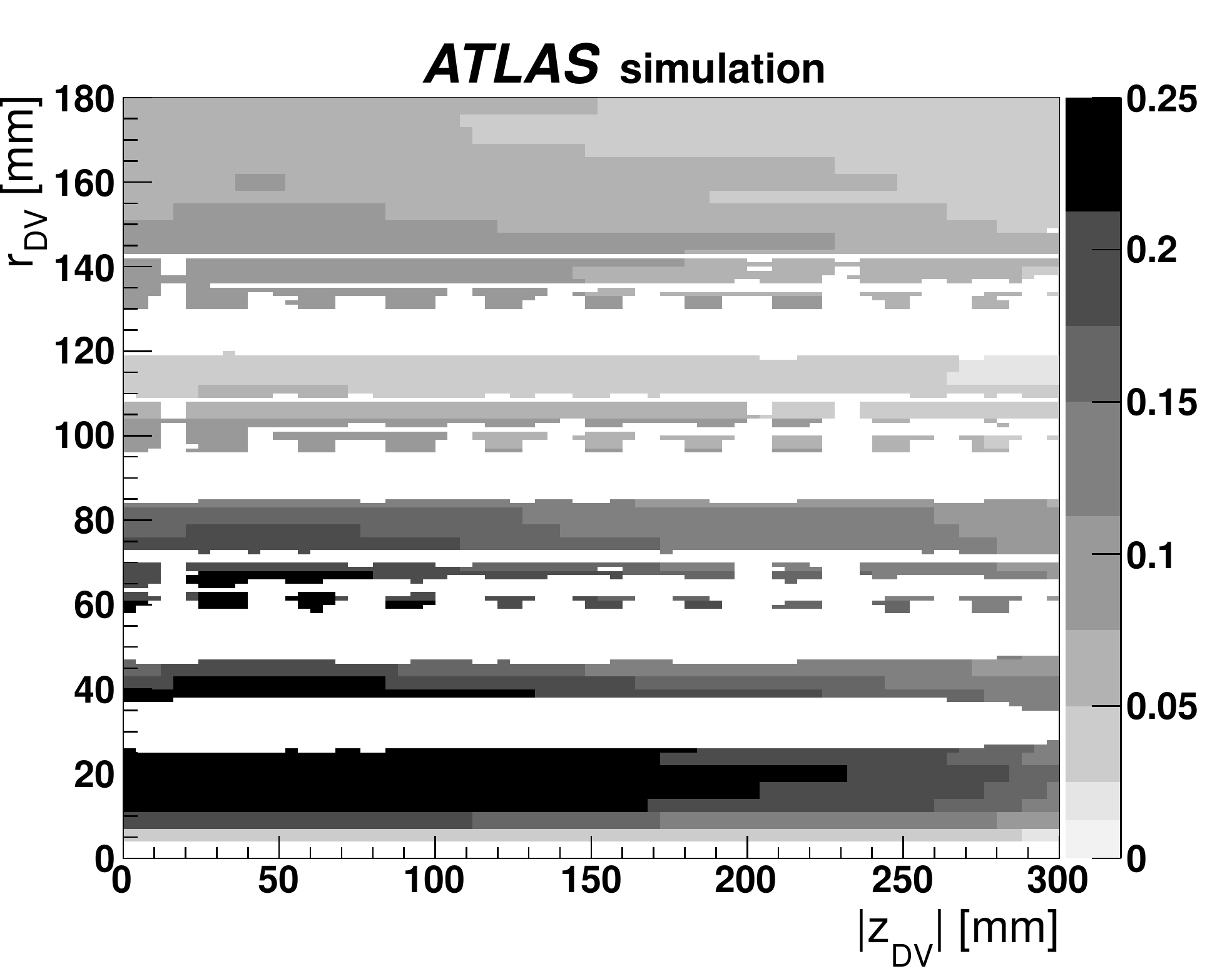}
   \caption{\label{fig:eff-2d} The efficiency as a function of \rdv\ 
     and \zdv\ for vertices in the
     signal MC sample~ML of Table~\ref{tab:sigMC}. High-density
     material regions that are vetoed are shown in white.
     The four major vetoed regions surround the beampipe wall,
     at radii of about $25-37~\mm$, and the three barrel pixel layers.
}
\end{figure}

We observe in MC studies that the vertex-efficiency dependence on \zdv\ is, to
a good accuracy, uncorrelated with its \rdv\ dependence.  Therefore, we
parametrize the efficiency function $\epsilon(\rdv, \zdv)$ as a
histogram in \rdv\ and a fourth-order polynomial in \zdv, with
parameters obtained from a fit to the MC. An example of the resulting
efficiency function, after including the effect of the material veto,
is shown in Fig.~\ref{fig:eff-2d}. 

\begin{figure}[!hbtp]
   \centering 
   \includegraphics[width=9cm]{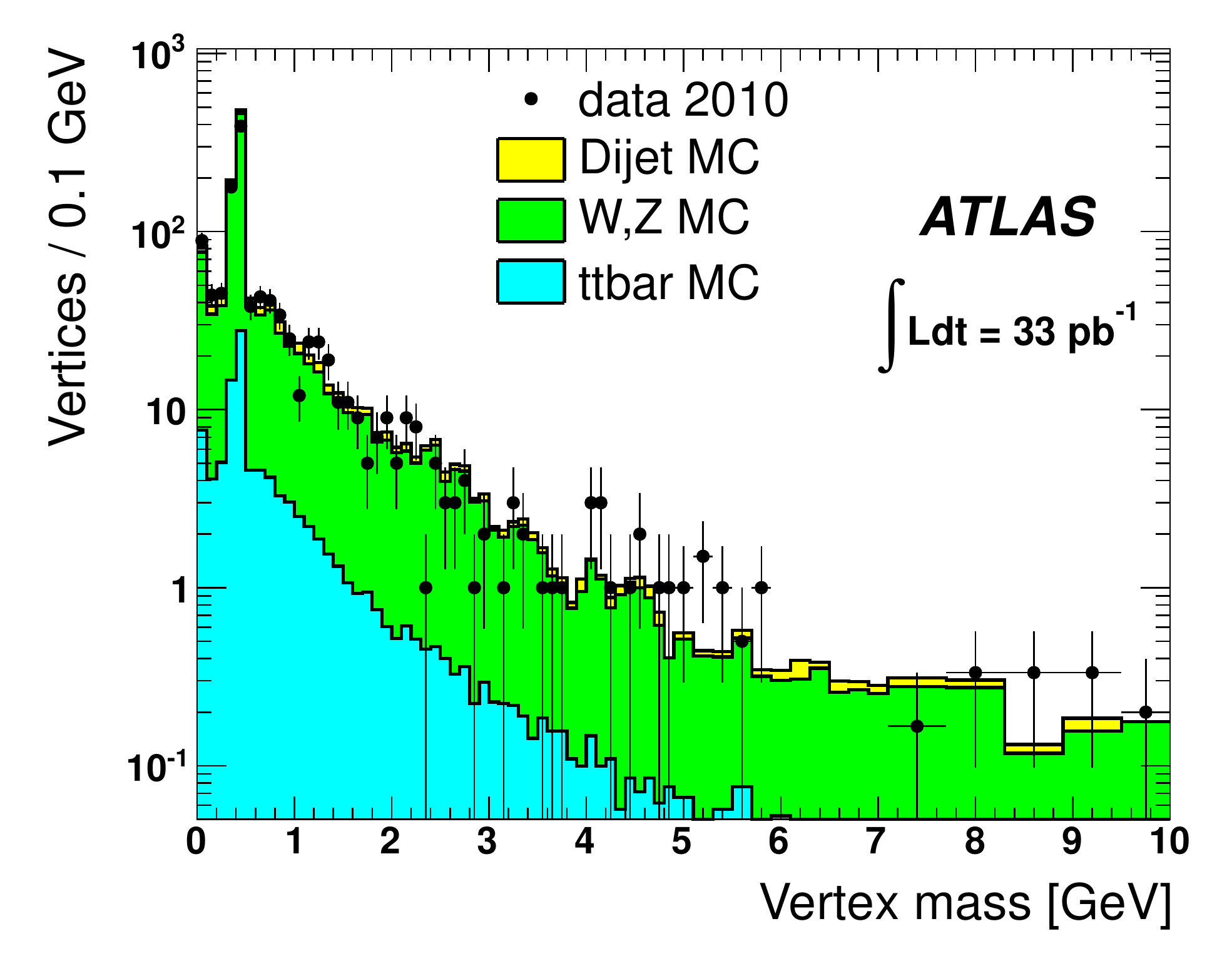} \\
   \includegraphics[width=9cm]{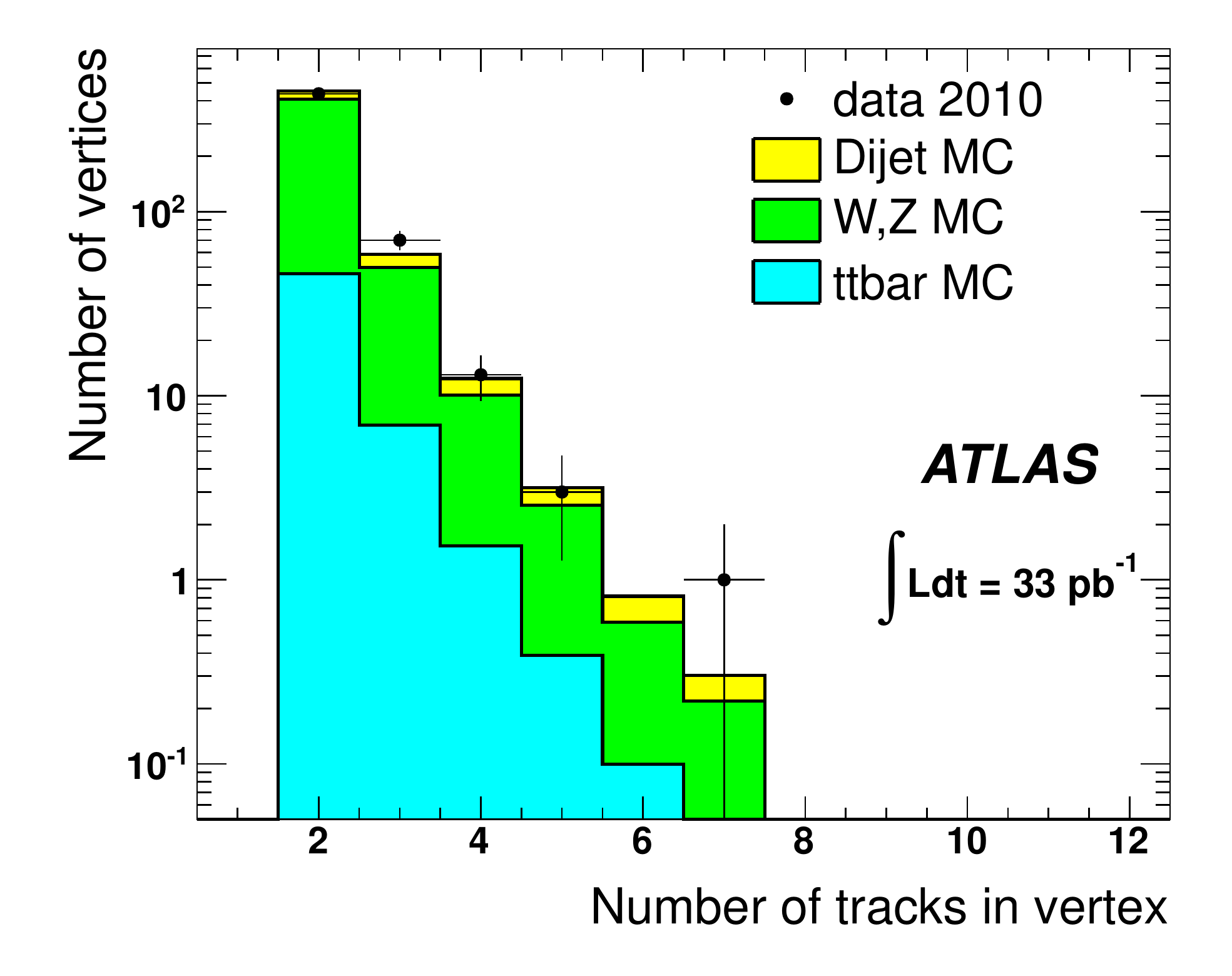} \\
   \includegraphics[width=9cm]{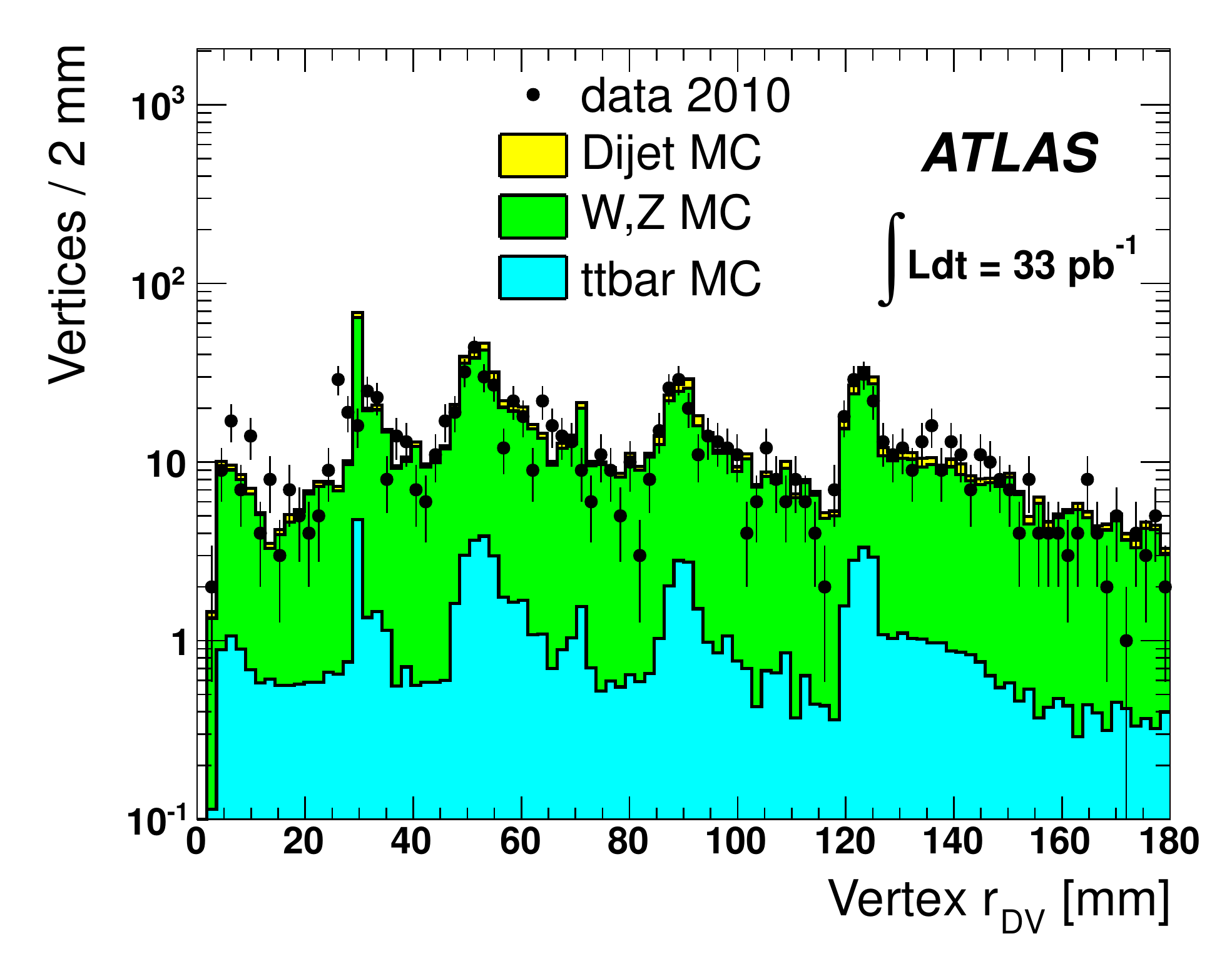} 
   \caption{\label{fig:data-mc} Comparison of the
\mdv\ (top), \ntdv\ (center), and \rdv\ (bottom)
distributions of data and MC events in the control region $\mdv <
10~\gev$. Other than the material veto and the 
$\ntdv \ge 4$ and $\mdv > 10~\gev$
requirements, all selection criteria are applied.  
In addition, the \ntdv\ and \rdv\ distributions include a veto
on \kshort\ decays.
The MC histograms are normalized to the integrated luminosity of the data,
with the MC cross-section given by PYTHIA~\cite{pythia}.}
\end{figure}

Fig.~\ref{fig:data-mc} shows comparisons between data and
background MC distributions of \mdv\ and \ntdv\ for selected events in
the control region, $\mdv < 10~\gev$ and before applying
the material veto.  Overall, good agreement
between data and MC is observed.  
As we show in Section~\ref{sec:bgd}, the level of background
contamination in our sample is estimated to be very small. Therefore,
small discrepancies between the data and MC distributions do not have
significant impact on the final results.
Additional studies comparing MC and control-sample data are described
in Section~\ref{sec:syst}.

\section{Background estimation}
\label{sec:bgd}

As we show later in this section, the probability for background
events to satisfy all the selection criteria is extremely
low. Therefore, in order to obtain enough events for background
estimation, we study the efficiency of the background to satisfy the
muon-selection criteria separately from the efficiency to satisfy the
other selection criteria. We then combine the results assuming that the
two efficiencies are uncorrelated.

We use the background MC samples (see Section~\ref{sec:data}) to
estimate the number of data events of each background type that are
expected to satisfy the selection criteria, without applying
any trigger requirements or the muon-selection criteria. 
Multiplying this number by
the probability for each MC event type to satisfy the muon-trigger and 
the offline muon-selection
criteria yields the expected background for each sample.
The $W^-\to\mu^-\bar \nu_\mu$ sample yields no selected vertices, but
has high efficiency for satisfying the muon requirements. As a result,
for this background we find the highest upper limit of all the other 
samples. Given 0 observed $W^-\to\mu^-\bar \nu_\mu$ MC events and the
luminosities of the data and of the MC sample,
we find the expected $W^-\to\mu^-\bar \nu_\mu$  background yield to be 
$N_{\rm bgd} < 0.03$ events at 90\% confidence level.
The expected background yield from $Z$, $t\bar t$, and
dijet events is at least an order of magnitude smaller.

We validate the use of MC to estimate the background by comparing
displaced-vertex yields in a sample of non-diffractive MC events and
data collected with minimum-bias triggers. For this study, we select
vertices with $\mdv<10~\gev$ and reject vertices with \mdv\
corresponding to \kshort\ or $\Lambda^0$ decays 
or to photon conversions, in order to increase the purity of 
material-interaction vertices with high position resolution.
From MC, we determine $\purity(\rdv)$, the radius-dependent fraction
of vertices that are due to particle interactions with material. This
fraction is close to unity in detector material and much smaller than
unity in gap regions between material layers, which are filled with N$_2$ 
gas.
Using $\purity(\rdv)$ and
the number of 2-track vertices in a pixel layer and in the adjacent
gap, we determine an effective pixel-layer-to-gas mass-density ratio
$\rho$. From $\rho$, $\purity(\rdv)$, and the number of $\ntdv>2$
vertices seen in each pixel layer, we predict the expected number of
such vertices in the adjacent gap. Comparing this with the number
of vertices actually observed, we find the prediction to be accurate
within expected statistical variations in both MC and data.

As a further cross-check of the estimated background level
in the muon-trigger sample, we study a control sample of events
selected with jet-based triggers and which fail the $\pt>45$~\gev\
muon trigger. These events are required to satisfy all the 
selection criteria, except the muon-selection 
and $\mdv>10~\gev$ requirements. 
We denote with $N^{\rm jet}_{\rm pass}$ the
number of control-sample events containing a vertex 
that satisfies the $\mdv>10~\gev$
requirement.
Then an estimate for the $\mdv>10~\gev$ background yield in the
muon-trigger sample is
$ N_{\rm bgd} = N^{\rm jet}_{\rm pass}
  N^{\mu}_{\rm fail} / N^{\rm jet}_{\rm fail} = 0.003 \pm 0.002$
events, 
where $N^{\mu}_{\rm fail}=3$ ($N^{\rm jet}_{\rm fail}=4170$) is the
number of selected muon-trigger 
(control-sample) events with no vertices that pass the
$\mdv>10~\gev$ requirement, and  $N^{\rm jet}_{\rm pass} = 4$.
We perform variations of this estimate to account for some differences
in the track-momentum spectrum and in \ntdv\ between the muon-trigger
and control-sample events, and obtain consistent results.

We observe that the data-driven background-yield estimate is consistent 
with the MC-based upper limit of $N_{\rm
bgd} < 0.03$ events. In
Section~\ref{sec:results} we conservatively use the MC-based upper limit
when calculating upper limits on the signal yield.

\section{Systematic uncertainties}
\label{sec:syst}

We study several sources of systematic uncertainties on the signal
reconstruction efficiency. These include uncertainties due to the trigger
efficiency, the $|d_0|$ dependence of muon- and hadron-reconstruction
efficiencies, the performance of the vertex-reconstruction algorithm,
the uncertainty on the integrated luminosity, and the effect of 
multiple $pp$ interactions per LHC bunch crossing. These studies and 
the resulting systematic uncertainties are briefly described below.
We estimate that systematic uncertainties associated with the
background estimate have a negligible impact on the limits obtained in
Section~\ref{sec:results}, since the conservatively estimated
background yield is far less than one event.

The efficiency of the trigger used in this analysis has been studied
using $Z^0\to\mu^+\mu^-$ events.  Based on
that study, we correct the MC-evaluated efficiency by an overall factor
of 0.91, and assign to it a relative uncertainty of 0.043.
Although $Z^0$ bosons are produced
at the interaction point, the trigger uses only MS hits, so it is
not strongly dependent on the true $|d_0|$ value. We find good agreement
between MC and data in terms of the dependence of the efficiency on the $|d_0|$
measured by the MS.

We test the MC simulation of the $|d_0|$ dependence of the offline
muon-reconstruction efficiency using cosmic-ray muons found 
in $pp$-collision events.
A downward-going cosmic-ray muon is reconstructed by the tracking
algorithm as two separate tracks, one above the interaction point and
one below.  To identify cosmic-ray muons, we select muon-candidate
pairs that are back-to-back to within 0.03 in $\phi$ and 0.01 in
$\eta$, and reject beam-produced muons by requiring $|d_0|>2~\mm$ for
both candidates.
The difference $\Delta\pt$ between the $\pt$ of the upper
candidate and that of the lower candidate peaks at around
6~\gev, due to muon energy loss in the calorimeters, and the fact
that for high-$\pt$ muons the $\pt$ measurement is dominated by the MS.
From the distribution of $\Delta\pt$ we find 
that the purity of the cosmic-ray sample is greater than 95\%.

\begin{figure}[!hbtp]
\begin{center}
  \includegraphics[width=9cm]{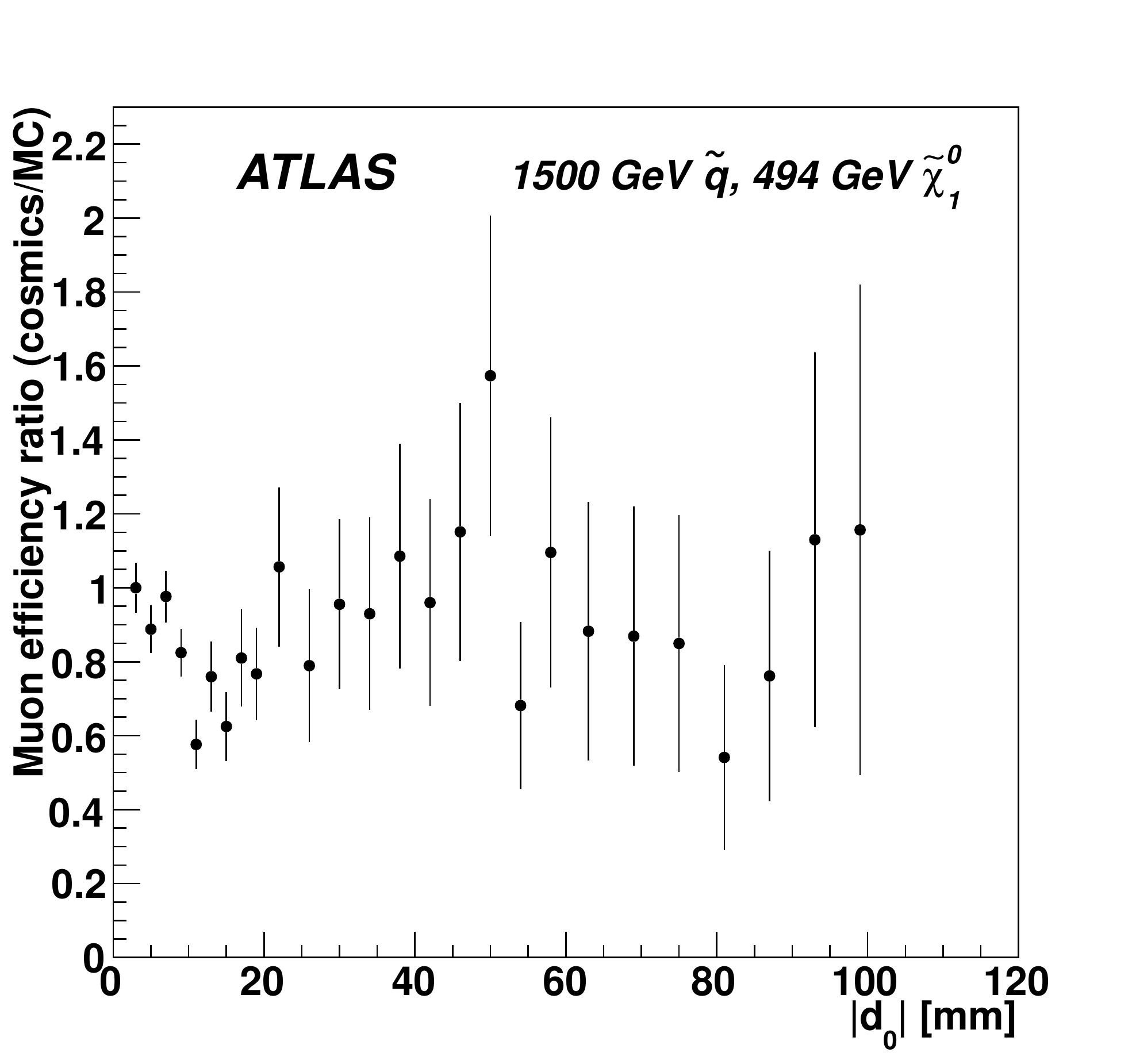}
  \caption{\label{fig:cosmic-eff} The ratio between the $|d_0|$
  distribution of cosmic-ray muons and the $|d_0|$-dependence of the
  signal MC muon reconstruction efficiency,
  shown as an example for sample~HH of Table~\ref{tab:sigMC}.
  The ratio is normalized to unity in the range $2 < |d_0| < 4~{\mm}$.
}
\end{center}
\end{figure}

Since the true-$|d_0|$ distribution for cosmic-ray
muons is flat over the small $|d_0|$ range used in the
analysis, as verified by cosmic-ray simulation,
the measured $|d_0|$ distribution of cosmic-ray muons
yields the $|d_0|$ dependence of the efficiency. 
Fig.~\ref{fig:cosmic-eff} shows the ratio between the number of cosmic
muons and the MC muon efficiency as a function of $|d_0|$, normalized to
unity in the range $2 < |d_0| < 4~\mm$.  The ratio shown 
is for sample~HH of Table~\ref{tab:sigMC}, but is similar for the
other MC samples.
All deviations from unity appearing in Fig.~\ref{fig:cosmic-eff} (and
in the corresponding distributions for the other signal samples) are
taken into account in the calculation of the signal efficiency 
in each $d_0$ bin, and are
reflected in Figs.~\ref{fig:eff} and~\ref{fig:eff-2d}.  
The statistical uncertainties shown in Fig.~\ref{fig:cosmic-eff} are
propagated to uncertainties of between 3.5\% and 8\% on the signal 
reconstruction efficiency, depending on the signal sample.

We estimate a systematic uncertainty on the inner-detector track
reconstruction efficiency as a function of the radial position of the
vertex by comparing the \rdv\ distributions of \kshort\ mesons in
minimum-bias MC and in data events collected with a minimum-bias
trigger. We find that the \kshort\ momentum distributions in data and
MC are in good agreement, and hence expect the \rdv\ distributions to
agree as well.
To test this, we compute the ratio $R_{\kshort} \equiv \kappa_{\rm data}(\rdv)/
\kappa_{\rm MC}(\rdv)$, where $\kappa_{\rm i}(\rdv)$ is the ratio
between the number of \kshort\ mesons reconstructed in sample $i$ in a
radial range around \rdv\ and the number reconstructed inside the
beampipe. 
Based on the largest deviation of $R_{\kshort}$ from unity, we
determine a systematic uncertainty of 3\% for tracks with $|\eta|<1$
and 4.3\% for tracks with $|\eta|>1$. This track-reconstruction
uncertainty is propagated to an uncertainty on the
vertex-reconstruction efficiency by random removal of tracks from the
signal MC samples prior to vertex reconstruction and generation of
alternative versions of the efficiency function $\epsilon(\rdv,
\zdv)$.

A data-MC difference in the vertex-reconstruction efficiency is
difficult to distinguish from an effect due to tracking-reconstruction
efficiency. A vertex-reconstruction efficiency difference is likely to
lead to differences in the vertex-fit $\chi^2/{\rm DOF}$. 
Comparing the fraction of vertices for which the $\chi^2/{\rm DOF}$
is below 2.5 in data and MC, we estimate a systematic uncertainty 
of 0.7\%.

We take the systematic uncertainty on the luminosity to be 3.4\%~\cite{lumi}.

The impact of multiple $pp$ interactions per bunch crossing on the
level of background is studied with jet-trigger events with multiple
primary vertices, and is determined to be negligible.  The impact on
signal efficiency, studied with the $\left< \mu \right> = 5$
signal MC sample, is negligible as well.  Therefore, no systematic
uncertainties are assigned due to multiple interactions.

Propagation of the systematic uncertainties to the final results of the 
analysis is described in the following section.

\section{Results}
\label{sec:results}

\begin{figure}[!hbtp]
\begin{center}
  \includegraphics[width=9cm]{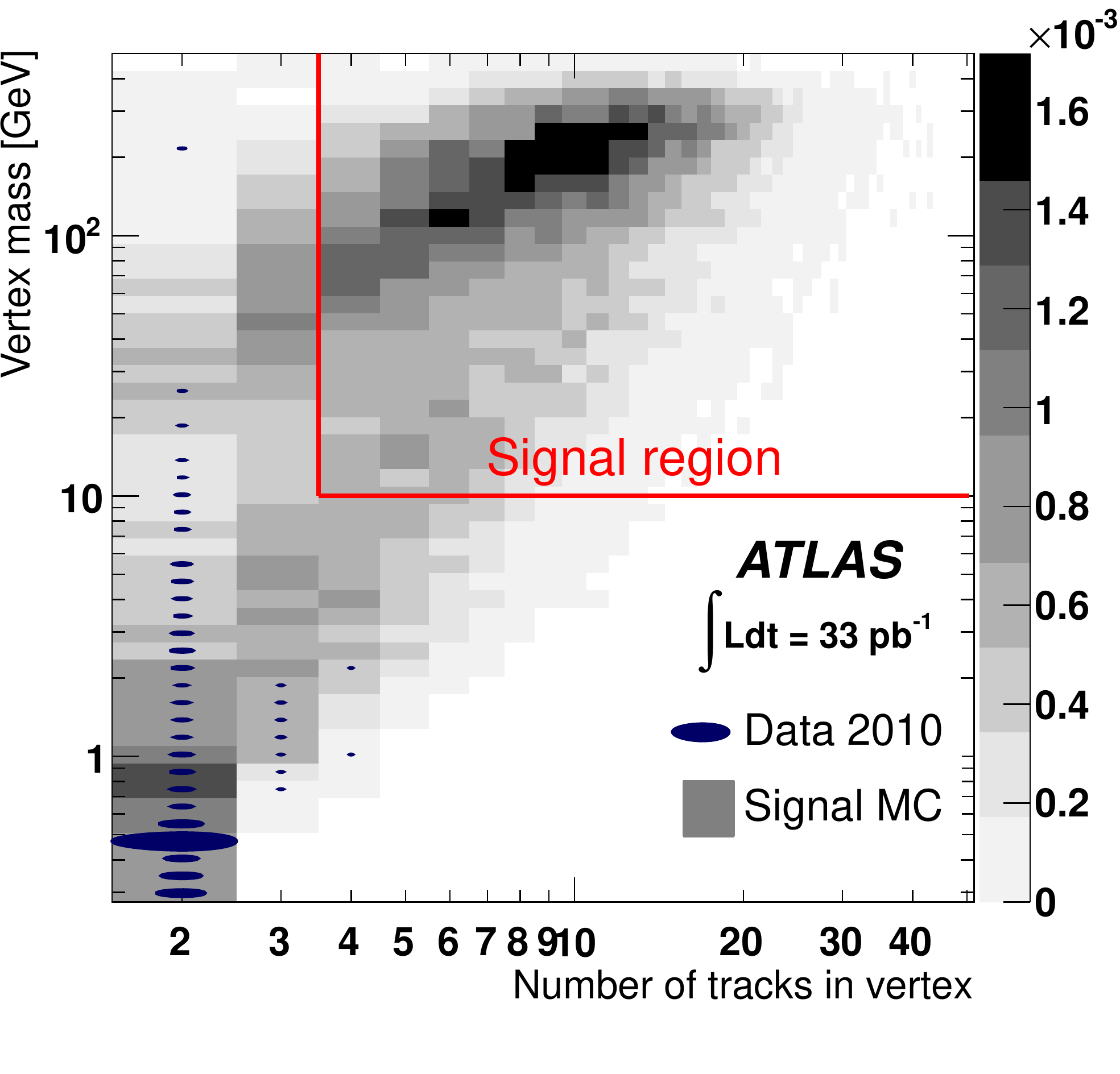}
  \caption{\label{fig:massVsMult} Vertex mass (\mdv) vs. vertex track
  multiplicity (\ntdv) for displaced vertices in events that pass the 
  selection requirements except the \mdv\ and \ntdv\ requirements,
  which are not applied.  Shaded bins show the distribution for the
  signal MC MH sample (see Table~\ref{tab:sigMC}), and data are shown as
  filled ellipses, with the area of the ellipse proportional to the
  number of vertices in the corresponding bin.  The figure contains 487
  data vertices, of which 251 are in the bin corresponding to \kshort\
  decays.}
\end{center}
\end{figure}

Figure~\ref{fig:massVsMult} shows the distribution of 
\mdv\ vs. \ntdv\ for vertices in the selected data events,
including vertices that fail the requirements on \mdv\ and \ntdv,
overlaid with the signal distribution for the MH sample.
We observe no events that satisfy all the selection criteria.

Based on this null observation, 
we set upper limits on the
supersymmetry production cross-section $\sigma$ times the branching
fraction $\cal B$ of the complete simulated signal decay chain for different
combinations of squark and neutralino masses and for different values
of $c\tau$, where $\tau$ is the neutralino lifetime.

The limits are determined in the following way.  For each value of
$c\tau$, we use the two-dimensional rapidity-vs.-velocity distribution
of the generated neutralinos in each signal MC sample to produce a
distribution of DV positions with respect to the PV.  
This distribution is convolved with a Gaussian representing the $z$
distribution of the PV, and then multiplied by the 2-dimensional
efficiency map for vertices in that signal MC sample, obtaining the
expected distribution of \rdv\ vs. \zdv.
This distribution is generated separately for two cases. In the
first case the reconstructed DV and muon originate from the same
neutralino, and in the second they originate from different
neutralinos. This allows us to correctly 
account for the muon-reconstruction efficiency 
for the desired value of $c\tau$, despite the fact that the signal MC
is produced with a different lifetime, $c\taumc$.
Integrating over the \rdv\ vs. \zdv\ distributions, we obtain the
total efficiency for reconstructing at least one vertex and one muon
in the event given our selection criteria and the value of $c\tau$.
From the efficiency and luminosity, we obtain the expected average
signal-event yield for any value of the signal production
cross-section.
The expected background yield is taken to be zero with a conservative
uncertainty of 0.03 events, which is the 90\% CL upper limit on the
background (see Section~\ref{sec:bgd}).
The upper limit on $\sigma {\cal B}$ is then calculated using the $CL_s$
method~\cite{CLs}, where signal-only and signal-plus-background
$p$-values are evaluated using pseudo-experiments generated from 
distributions based on counting statistics. The uncertainties on
luminosity, efficiency, and background are treated as nuisance 
parameters.

The systematic uncertainty on the track-reconstruction efficiency is
taken into account in the limit calculation by use of the alternative
efficiency functions described in Section~\ref{sec:syst}. All
other efficiency systematic uncertainties are used when converting
the limit on the number of signal events to the limit on $\sigma {\cal B}$.

\begin{figure}[!hbtp]
\begin{center}
  \includegraphics[width=9cm]{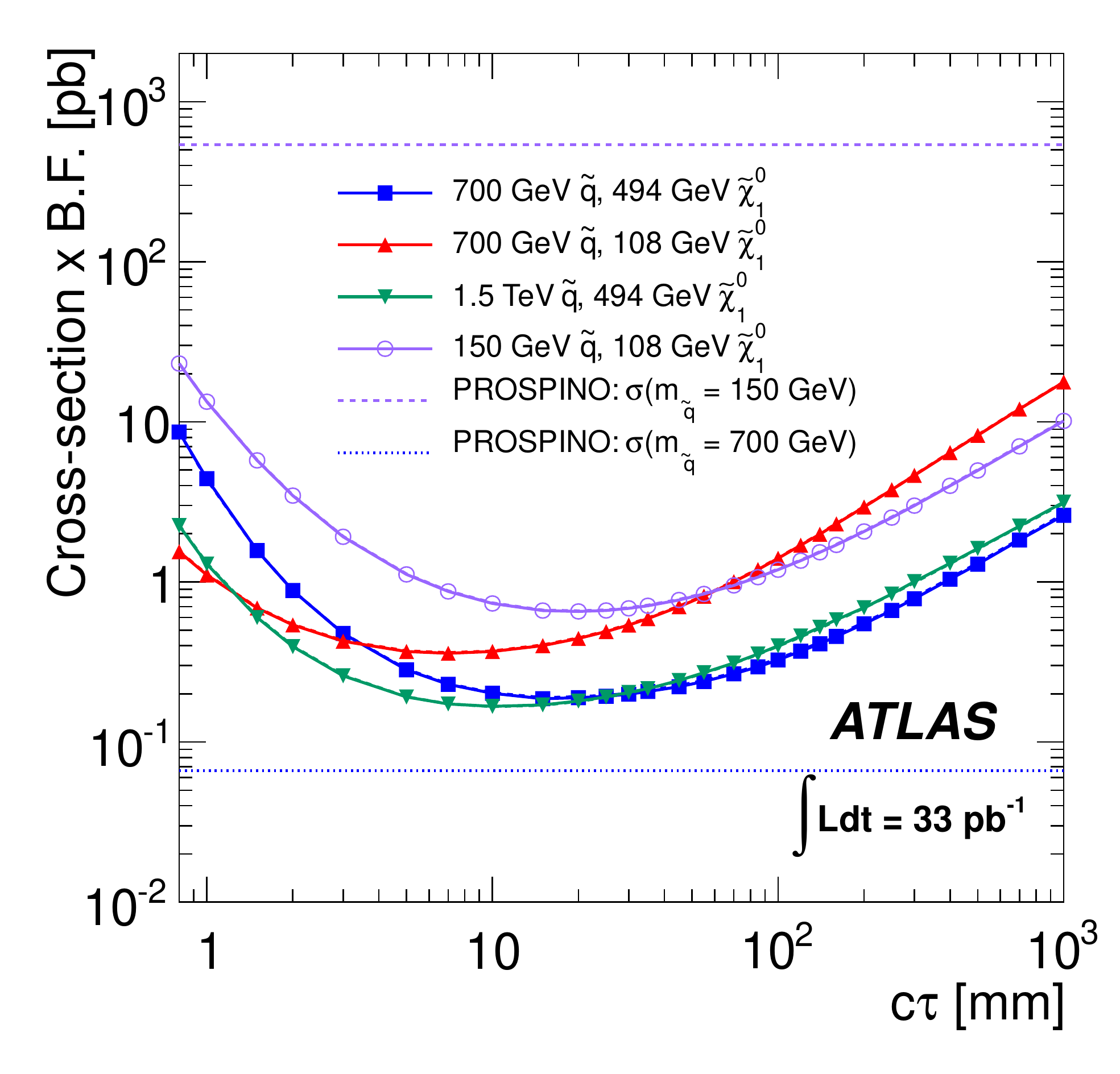}
  \caption{\label{fig:limits} Upper limits at 95\% CL on the production
  cross-section times branching fraction vs. the neutralino lifetime
  times the speed of light for different combinations of squark and
  neutralino masses, based on the observation of zero events 
  satisfying all criteria in a
  \lumi\ data sample. The horizontal lines show the cross-sections
  calculated from PROSPINO for squark masses of 700~\gev\ and 150~\gev.
}
\end{center}
\end{figure}

The resulting limits are shown in Figure~\ref{fig:limits},
with the PROSPINO-calculated cross-sections for squark masses of 150~\gev\
and 700~\gev.
Since no background is
expected, the expected and observed limits are indistinguishable.
In addition, based on the observation of no signal events in a data
sample of $\lumi$, we set a 95\% confidence-level upper limit of
0.09~pb on the cross-section times the detector acceptance times the
reconstruction efficiency for any signal vertex.

\section{Summary and conclusions}
\label{sec:conclusions}

We have searched for new, heavy particles with lifetimes and
velocities such that they decay at radial distances between 4~\mm\ and
180~\mm\ from the $pp$ interaction point, in association with a
high-transverse-momentum muon.
Fewer than 0.03 background events are expected in the data sample of \lumi,
and no events are observed. We present limits on the product of
di-squark production cross-section and decay-chain branching fraction
in a SUGRA scenario where the lightest neutralino produced in the
primary-squark decay undergoes \RP-violating decay into a muon and two
quarks. 
Limits are reported as a function of the neutralino lifetime and for a
range of neutralino masses and velocities, which are the factors with
greatest impact on the limit. Limits for a variety of other models can
thus be approximated from our results, based on the neutralino mass
and velocity distribution in a given model.


\section{Acknowledgements}

We thank CERN for the very successful operation of the LHC, as well as the
support staff from our institutions without whom ATLAS could not be
operated efficiently.

We acknowledge the support of ANPCyT, Argentina; YerPhI, Armenia; ARC,
Australia; BMWF, Austria; ANAS, Azerbaijan; SSTC, Belarus; CNPq and FAPESP,
Brazil; NSERC, NRC and CFI, Canada; CERN; CONICYT, Chile; CAS, MOST and
NSFC, China; COLCIENCIAS, Colombia; MSMT CR, MPO CR and VSC CR, Czech
Republic; DNRF, DNSRC and Lundbeck Foundation, Denmark; ARTEMIS, European
Union; IN2P3-CNRS, CEA-DSM/IRFU, France; GNAS, Georgia; BMBF, DFG, HGF, MPG
and AvH Foundation, Germany; GSRT, Greece; ISF, MINERVA, GIF, DIP and
Benoziyo Center, Israel; INFN, Italy; MEXT and JSPS, Japan; CNRST, Morocco;
FOM and NWO, Netherlands; RCN, Norway; MNiSW, Poland; GRICES and FCT,
Portugal; MERYS (MECTS), Romania; MES of Russia and ROSATOM, Russian
Federation; JINR; MSTD, Serbia; MSSR, Slovakia; ARRS and MVZT, Slovenia;
DST/NRF, South Africa; MICINN, Spain; SRC and Wallenberg Foundation,
Sweden; SER, SNSF and Cantons of Bern and Geneva, Switzerland; NSC, Taiwan;
TAEK, Turkey; STFC, the Royal Society and Leverhulme Trust, United Kingdom;
DOE and NSF, United States of America.

The crucial computing support from all WLCG partners is acknowledged
gratefully, in particular from CERN and the ATLAS Tier-1 facilities at
TRIUMF (Canada), NDGF (Denmark, Norway, Sweden), CC-IN2P3 (France),
KIT/GridKA (Germany), INFN-CNAF (Italy), NL-T1 (Netherlands), PIC (Spain),
ASGC (Taiwan), RAL (UK) and BNL (USA) and in the Tier-2 facilities
worldwide.

\bibliographystyle{atlasnote}
\bibliography{DV-pap}

\providecommand{\href}[2]{#2}\begingroup\raggedright\begin{thebibliography}{10}


\bibitem{Allanach:2006st} 
  B.~C.~Allanach {\it et al.},
  Phys.\ Rev.\  D {\bf 75}, 035002 (2007)
  [arXiv:hep-ph/0609263].


\bibitem{Dimopoulos:1996vz}
  S.~Dimopoulos, M.~Dine, S.~Raby and S.~D.~Thomas,
  Phys.\ Rev.\ Lett.\  {\bf 76}, 3494 (1996)
  [arXiv:hep-ph/9601367].


\bibitem{Hewett:2004nw}
  J.~L.~Hewett, B.~Lillie, M.~Masip and T.~G.~Rizzo,
  JHEP {\bf 0409}, 070 (2004)
  [arXiv:hep-ph/0408248].


\bibitem{Strassler:2006im}
  M.~J.~Strassler and K.~M.~Zurek,
  Phys.\ Lett.\  B {\bf 651}, 374 (2007)
  [arXiv:hep-ph/0604261].


\bibitem{Schuster:2009au}
  P.~Schuster, N.~Toro and I.~Yavin,
  Phys.\ Rev.\  D {\bf 81}, 016002 (2010)
  [arXiv:0910.1602 [hep-ph]].


\bibitem{Fan:2011yu}
  J.~Fan, M.~Reece, J.~T.~Ruderman,
    [arXiv:1105.5135 [hep-ph]].


\bibitem{Zur:2008zg}
  B.~K.~Zur, L.~Mazzucato and Y.~Oz,
  JHEP {\bf 0810} (2008) 099
  [arXiv:0807.4543 [hep-ph]].


\bibitem{D0:2006}
  The D0 Collaboration, V.~M.~Abazov {\it et al.},
  Phys.\ Rev.\ Lett.\  {\bf 97}, 161802 (2006)
  [arXiv:hep-ex/0607028].

\bibitem{D0:2009}
  The D0 Collaboration, V.~M.~Abazov {\it et al.},
  Phys.\ Rev.\ Lett.\  {\bf 103}, 071801 (2009)
  [arXiv:0906.1787 [hep-ex]].
  
\bibitem{ATLAS:detector2008}
  The ATLAS Collaboration, 
  JINST {\bf 3}, S08003 (2008).

\bibitem{pythia}
T.~Sj{\"o}strand, S.~Mrenna, and P.~Z.~Skands,
  JHEP {\bf 0605}, 026 (2006)
  [hep-ph/0603175].

\bibitem{MC_NLO}
S.~Frixione and B.~R.~Webber,
JHEP {\bf 06} 029 (2002).

\bibitem{Beenakker:1996ed}
  W.~Beenakker, R.~Hopker, M.~Spira,
[hep-ph/9611232].


\bibitem{geant4}
The GEANT4 Collaboration, S.~Agostinelli {\it et al.},
Nucl. Instrum. Meth. A {\bf 506} (2003).

\bibitem{ATLASgeant4}
  The ATLAS Collaboration,
  Eur.\ Phys.\ J.\  {\bf C70}, 823-874 (2010).
  [arXiv:1005.4568 [physics.ins-det]].


\bibitem{vtxNote} 
The ATLAS Collaboration,
  [arXiv:1110.6191 [hep-ex]].




\bibitem{lumi}
  G.~Aad {\it et al.} [ ATLAS Collaboration ],
  Eur.\ Phys.\ J.\  {\bf C71}, 1630 (2011).
  [arXiv:1101.2185 [hep-ex]].


\bibitem{CLs}
  A.~L.~Read,
  J.\ Phys.\ {\bf G28}, 2693-2704 (2002).
  



\end{thebibliography}\endgroup


\clearpage
\onecolumn
\begin{flushleft}
{\Large The ATLAS Collaboration}

\bigskip

G.~Aad$^{\rm 48}$,
B.~Abbott$^{\rm 111}$,
J.~Abdallah$^{\rm 11}$,
A.A.~Abdelalim$^{\rm 49}$,
A.~Abdesselam$^{\rm 118}$,
O.~Abdinov$^{\rm 10}$,
B.~Abi$^{\rm 112}$,
M.~Abolins$^{\rm 88}$,
H.~Abramowicz$^{\rm 153}$,
H.~Abreu$^{\rm 115}$
% ...
\bigskip

% ...
$^{10}$ Institute of Physics, Azerbaijan Academy of Sciences, Baku, Azerbaijan\\
$^{11}$ Institut de F\'isica d'Altes Energies and Departament de F\'isica de la Universitat Aut\`onoma  de Barcelona and ICREA, Barcelona, Spain\\
% ...
$^{48}$ Fakult\"{a}t f\"{u}r Mathematik und Physik, Albert-Ludwigs-Universit\"{a}t, Freiburg i.Br., Germany\\
$^{49}$ Section de Physique, Universit\'e de Gen\`eve, Geneva, Switzerland\\
% ...
$^{88}$ Department of Physics and Astronomy, Michigan State University, East Lansing MI, United States of America\\
% ...
$^{111}$ Homer L. Dodge Department of Physics and Astronomy, University of Oklahoma, Norman OK, United States of America\\
$^{112}$ Department of Physics, Oklahoma State University, Stillwater OK, United States of America\\
% ...
$^{115}$ LAL, Univ. Paris-Sud and CNRS/IN2P3, Orsay, France\\
% ...
$^{118}$ Department of Physics, Oxford University, Oxford, United Kingdom\\
% ...
$^{153}$ Raymond and Beverly Sackler School of Physics and Astronomy, Tel Aviv University, Tel Aviv, Israel\\
% ...
\end{flushleft}


\end{document}